# Common trends in the epidemic of Covid-19 disease

**Milad Radiom, Jean-François Berret**

*Université de Paris, CNRS, Matière et systèmes complexes, 75013 Paris, France*

Correspondence: milad.radiom@univ-paris-diderot.fr, jean-francois.berret@univ-paris-diderot.fr

**Abstract.** The discovery of SARS-CoV-2, the responsible virus for the Covid-19 epidemic, has sparked a global health concern with many countries affected. Developing models that can interpret the epidemic and give common trend parameters are useful for prediction purposes by other countries that are at an earlier phase of the epidemic; it is also useful for future planning against viral respiratory diseases. One model is developed to interpret the fast-growth phase of the epidemic and another model for an interpretation of the entire data set. Both models agree reasonably with the data. It is shown by the first model that during the fast phase, the number of new infected cases depends on the total number of cases by a power-law relation with a scaling exponent equal to 0.82. The second model gives a duplication time in the range $1 - 3$ days early in the start of the epidemic, and another parameter ($\alpha = 0.1 - 0.5$) that deviates the progress of the epidemic from an exponential growth. Our models may be used for data interpretation and for guiding predictions regarding this disease, *e.g.* the onset of the maximum in the number of new cases.

**Abbreviation:** SARS = Severe Acute Respiratory Syndrome; CoV = Corona Virus

**Graphical Abstract:**

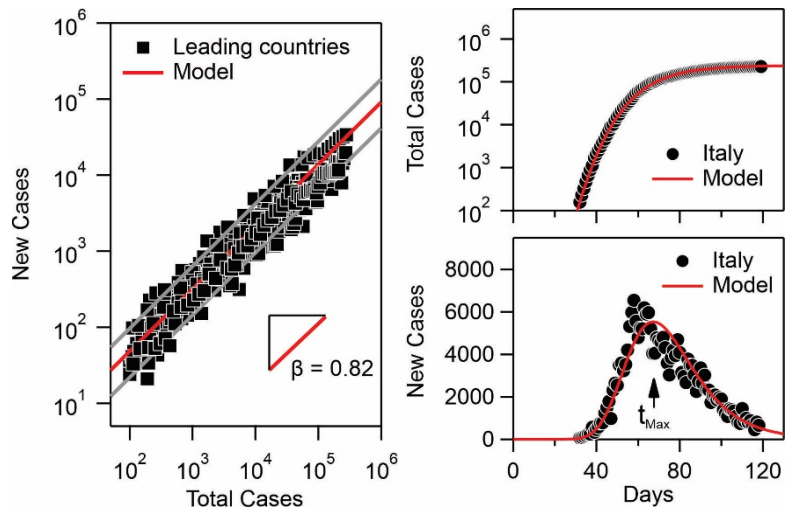

## 1. Introduction

An outbreak of an infectious respiratory disease emerged in the city of Wuhan, in the Chinese province of Hubei, in December 2019. One month later, the discovery of the responsible virus, SARS-CoV-2, was announced [1]. The respiratory disease, which is termed Covid-19, is an ongoing global health concern with many countries affected so far. While a great momentum in the scientific community is to find therapeutics and vaccines against Covid-19 [2], it seems beneficial to develop *simple* models to understand the trends of the epidemic [3, 4]. Previous attempts have been made to this end. New models are generally adopted from the previous models of avian and swine flu epidemics [5, 6], and may be used to make predictions regarding the Covid-19 epidemic, *e.g.* to simulate spreading to neighboring regions from hot spots, and to evaluate the effectiveness of the mitigating measures [7-9]. Available models have generally estimated a human-to-human transmissibility of disease, which is described by the basic reproduction number, $R_o$. In particular, $R_o > 1$ is associated with an infection that will spread in a population, whereas $R_o < 1$ corresponds to an epidemic that will decrease. Estimations of $R_o$ for Covid-19 give values larger than 2 denoting a highly contagious virus [5, 10, 11]. Simpler models exist and have been used for example to compare between different SARS epidemics [12, 13].

This letter seeks to interpret the available data using a simple model making use of physically meaningful parameters. In particular, with respect to the several periods for an epidemic, the presented analysis focuses on the first period, before the so-called herd immunity period. In the herd immunity and in the absence of a vaccine, the virus stops expanding when 60% of a population is contaminated [14-16]. A Californian study has shown that the number of contaminated people in this region was 50 times more than the total number of reported cases [17]. Applying this factor to France's numbers, with a total case of about $1.8{\times}10^5$, this factor gives about 7.5 million contaminated cases. This number of cases is equivalent to 12% of the French population. Recently, another study stated that this number was about 5.7% on average and about 12% in the region Ile-de-France [18]. In general, epidemiological models focus on the number of new cases as a function of time. In the first period of the epidemic – sometimes described as the first "wave" – the number of new cases shows characteristic features: the initial exponential growth is followed by a maximum and then a decrease. The shape of this curve depends on various factors such as the basic reproduction number associated with the epidemic, environmental conditions or the behavior of the infected population, among others. Later the epidemic can evolve following different scenarios, which can be a steady and low background of new cases, or a second wave; the latter showing characteristics similar to those of the first wave. The available data and the model presented in this work are for the first period of the epidemic.

The focus of this letter is initially placed on the data that is available for China and South Korea because at the time of writing this letter these countries have *apparently* reached the end of the first period of their epidemic. The data from these countries together with a few others, such as Iran, Turkey, Italy, Spain, France, the United States, Russia, Germany, the United Kingdom and Brazil are considered to develop common trend parameters using a simple physical model. The developed models in this work may be used for prediction purposes by other countries or with respect to future respiratory viral infections.

## 2. Results and discussion

The data of temporal variation of total number of cases is gathered from an online source [19] from 22-02-2020 to 20-05-2020 and reproduced in Figs. 1a and 1b for China and South Korea, respectively. It is expected that the trends of the other countries during the pre-herd immunity period will be similar. In particular for China and South Korea, in the first period of the epidemic, three phases may be assigned to the temporal variation of the total number of cases data, whereby an initial fast-growing phase is followed by a slow-growing phase and then a plateau. These phases are respectively marked $f$ (fast), $s$ (slow) and $p$ (plateau) on Figs. 1a and 1b. These phases may be discerned from the apparent variations in the growth exponent of the total number of cases versus time. For example, in the $f$ phase of China, the apparent growth exponent $(1/n)(dn/dt)$, where $n$ is the total number of cases and $t$ the time, is about 0.1, while it is 0.006 and $< 0.001$ respectively in the $s$ phase and the $p$ phase. For South Korea, the apparent growth exponents are 0.137, 0.006 and 0.001 respectively during the $f$ phase, the $s$ phase, and the $p$ phase. Since the data are presented as a function of the number of days starting from 22-02-2020, a note should be made for the data-days prior to this date. This data is not available; however, it may be assumed that the temporal variation of the total number of cases prior to 22-02-2020 follows the behavior of the $f$ phase. The onset of the $s$ phase and the $p$ phase is about two to four weeks for both China and South Korea.

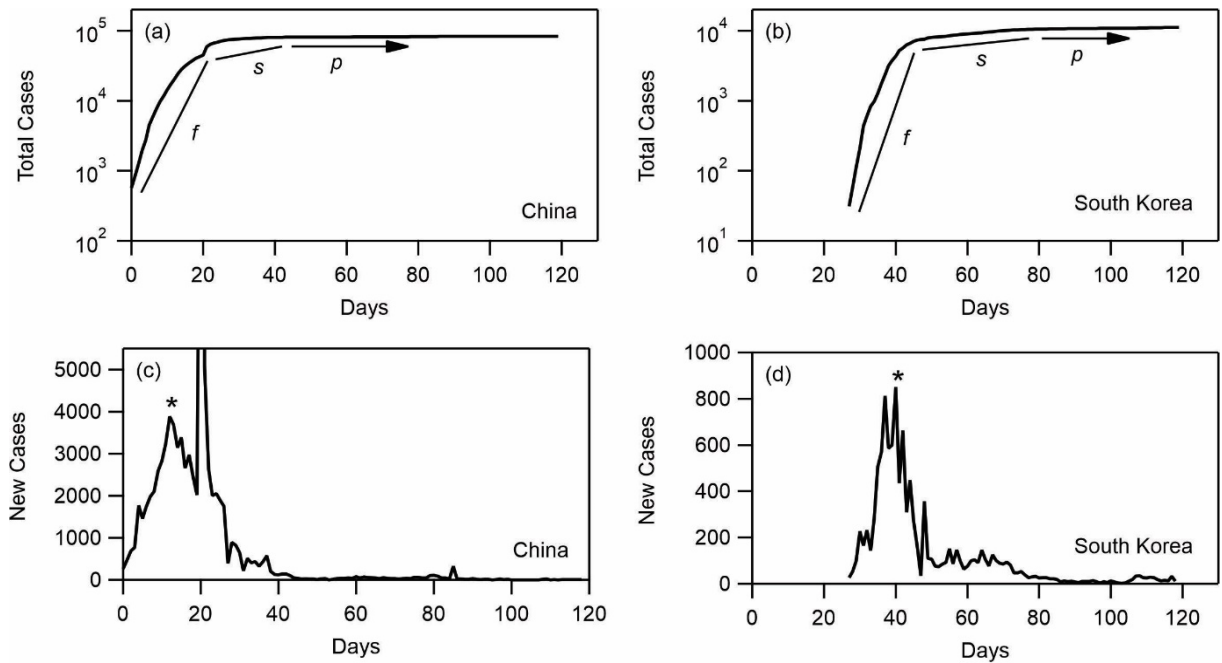

***Figure 1***: *(a, b) Total number of cases as a function of the number of days starting from 22-01-2020 for China (a) and South Korea (b). The fast-growing phase (f), slow-growing phase (s) and plateau phase (p) are marked. These phases may be discerned from the apparent variations in the growth exponent. (c, d) Number of new cases as a function of the number of days starting from 22-01-2020 for China (c) and South Korea (d). Peak in the number of new cases is marked with asterisks. All data are collected until 20-05-2020.*

The number of new cases is calculated from the previous data and presented in Figs. 1c and 1d respectively for China and South Korea. In particular, these data show peaks which are marked with asterisks on Figs. 1c and 1d. As mentioned in the introduction, a peak-shaped number of new cases is a common feature of epidemic outbreaks. The onset of the peak corresponds approximatively to the start of the $s$ phase discussed earlier [14].

The models presented hereafter in this work address the first period of the epidemic (before the herd immunity), when typically the proportion of the infected population is inferior to 10% [18]. One common trend is obtained by plotting the number of new cases as a function of the total number of cases in the $f$ phase, $i.e.$ prior to the onset of maximum in the number of new cases. Here we include the $f$ phase data from 12 different countries with leading total number of infected cases [19, 20]. As shown in Fig. 2, this data presentation leads to the expression:

$$dn = \lambda n^{\beta}, \tag{1}$$

where $dn$ is the number of new cases, $\lambda$ the prefactor and $\beta$ a scaling exponent. It is found that Eq. 1 can reasonably interpret the data with an exponent $\beta = 0.82$. The prefactor $\lambda$ is found to vary from 0.5 to 2.2 for all the investigated countries, including China, South Korea, Iran, Turkey, Italy, Spain, France, the United States, Russia, Germany, the United Kingdom, and Brazil. The high value of the exponent $\beta$ signifies the wide spread of the virus during the $f$ phase. It also shows that the number of new cases is strongly dependent on the total number of cases during the $f$ phase. However, as $\beta < 1$, the spread during this phase is sub-exponential. Similar plots for individual countries are included in the supplementary information, Fig. SI 1.

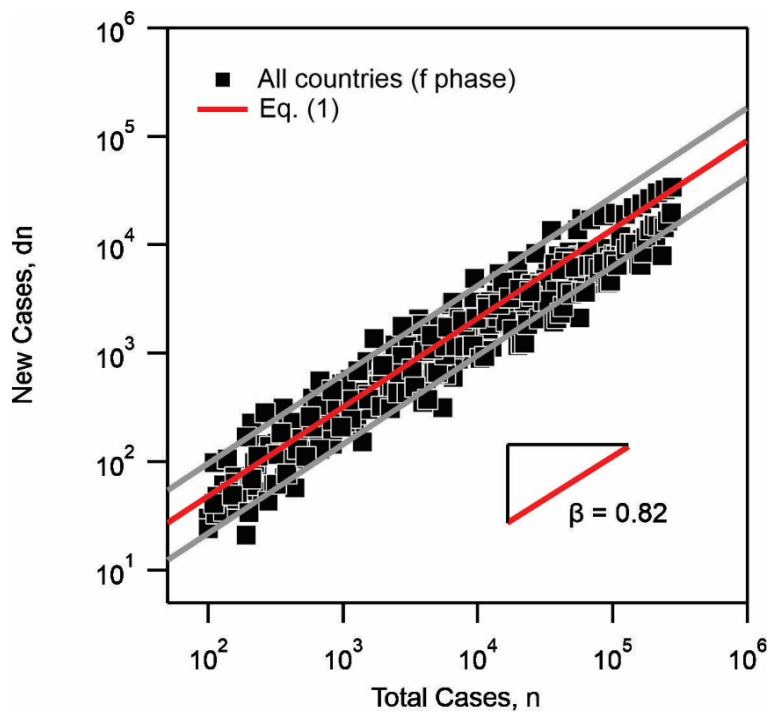

**Figure 2**: Number of new cases as a function of the total number of cases showing the power-law dependence of Eq. 1. The exponent $\beta = 0.82$ is found to interpret the data reasonably. The *prefactor $\lambda = 0.5$ gives the lower bound and $\lambda = 2.2$ the upper bound. Data from all countries (12 countries) are included [19, 20]. In supplementary information Fig. SI 1 we provide the figures for individual countries.*

Table 1 gives predictions for the number of new cases from the total number of cases during the *f* phase. The values are calculated for the exponent $\beta = 0.82$, and the prefactor $\lambda = 0.5$ (resp. $\lambda = 2.2$) to obtain a minimum (resp. maximum) value. The results show that by controlling the total number of cases at the start of the epidemic, significant reduction in the number of new cases can be obtained. It further shows that 1 to 3 orders of magnitude increase in the total number of cases in the *f* phase, results in about $7 - 288$ times higher number of new cases.

Table 1: Predictions for the number of new cases from the total number of cases during the *f* phase. Calculations are based on Eq. 1, for $\beta = 0.82$, , and $\lambda = 0.5$ to obtain a minimum and $\lambda = 2.2$ to obtain a maximum.

| Total cases, $n$ | Minimum new cases, $dn_{\min}$ | Maximum new cases, $dn_{\max}$ |
|---|---|---|
| 100 | 22 | 96 |
| 1000 | 144 | 634 |
| 10000 | 953 | 4192 |
| 100000 | 6295 | 27696 |

Next, a common trend for the temporal evolution of the total number of cases is sought. The following expression is used to interpret the latter data [21]:

$$n = 2^{t/\tau} \left[ \frac{1}{n_o^{\alpha}} + \frac{1}{n_{\infty}^{\alpha}} \left( 2^{\alpha t/\tau} - 1 \right) \right]^{-1/\alpha}, \qquad (2)$$

where $n_o$ and $n_{\infty}$ correspond to the total number of cases at the start and the end of epidemic respectively (*i.e.* the boundaries), $\tau$ is similar to the duplication time, and $\alpha$ is a parameter that controls the temporal variation of the data trend, strictly its deviation from an exponential growth which we find to occur early in each data set. A high value of $\alpha$ also corresponds to a fast transition from the *f* phase to the *s* phase and the *p* phase.

The form of Eq. 2 is inspired from the growth of eukaryotic cells in Petri dish, where the boundary conditions slow the initial exponential growth [21]. For cell culture, it was found that $\alpha = 2$ gave good results. Here, $\alpha$ is left as an adjustable parameter controlling the different measures taken by the states to mitigate the virus expansion, including social distancing and barrier gestures. Although the effects of social confinement are different from Petri dish boundary conditions, it is shown below that Eq. 2 can reasonably interpret the data. It is noted that other functions may be used for data interpretation [11, 14, 22, 23]; the preference in the use of Eq. 2 is its simplicity and physically meaningful parameters. For example as compared to Gumbel's model recently adopted to interpret the data, we show our model to give reasonable agreements over a longer period of the epidemic [23, 24]. It is noted however that cell division is fundamentally different from disease spreading in that, in the simplest form of argument, one cell divides to two related daughter cells while one individual can infect more than one related or non-related individuals. The applicability of Eq. 2 is in its form and should be not taken as a parallel between cell growth process and disease spreading process.

However, it is not possible to fit all available data to Eq. 2 as many countries are in their $f$ phase and a value of $n_\infty$ cannot be assigned without uncertainties. Data from the same selected countries as in Fig. 2 were then fitted to Eq. 2. These fits for China, South Korea, Iran, and Turkey are shown in Fig. 3. It is noted that fitting started from $n \geq 100$, and the value of $n_o$ was fixed. Thereby, the number of the fitting parameters was reduced to 3, $n_\infty$, $\tau$ and $\alpha$. From Fig. 3, it is evident that Eq. 2 provides a reasonable agreement with the data, especially considering occasional non-uniformities in the trend, *e.g.* in the case of South Korea's and Iran's data. These non-uniformities are best presented in the number of new cases trends. For example, South Korea's number of new cases data shows one peak at around day 40 and a second peak at around day 65. In Iran's data, two peaks at around day 50 and day 65 are observed while another peak is expected to occur beyond day 120. In this later case, the data could correspond to a sequence of three consecutive waves, each of them being potentially described with Eq. 2. The best fit parameters are presented in Table 2. One observes that the duplication time varies in the range $1 - 3$ days among these countries and the $\alpha$ parameter is in the range of $0.2 - 0.5$. China and Iran present similar duplication time equal to 1.7 days and 1.5 days respectively while their $\alpha$ parameter differ and are respectively 0.45 and 0.17. A higher $\alpha$ parameter in the case of China corresponds to this country's faster transition from the $f$ phase to the $s$ phase as compared to Iran. The transition from the $f$ phase to the $s$ phase has a reverse dependence on the duplication time. For example, for China and Turkey which have a similar $\alpha$ parameter, a shorter duplication time for China as compared to a longer one equal to 2.7 days for Turkey means that China's transition occurred over a shorter period of time.

It is noted that the duration of $f$ phase is much longer than the duration of the exponential growth incorporated in Eq. 2. Indeed, we find that most countries have a short time span of exponential growth which is deviated in a matter of a few days. Thereby the parameters associated with Eq. 1 resulting in the comparisons shown in Fig. 2, and those of Eq. 2 resulting in Figs. $3 - 5$ are not interpreting the same range of data. More precisely, $f$ phase is all data up to the maximum in the number of new cases, while the exponential growth in the data only lasts a few days.

Furthermore, the number of new cases is calculated from the total number of cases as discussed previously. Thereby, Fig. 3 further shows a comparison between the latter data and calculations from the derivative of Eq. 2 using the same fit parameters. The reasonable agreement obtained from the latter comparison further shows that Eq. 2 is adequate for data interpretation. Similarly, fits to the total of number of cases for Italy, Spain, France, and the United States, together with the reproduction of the number of new cases from data and from derivative of Eq. 2 are shown in Fig. 4. Figure 5 shows similar data for Russian, Germany, the United Kingdom and Brazil. The agreement is again reasonable. We note that for certain countries such as Italy, France, Germany, Russia and UK, the agreement is excellent both on the total number of cases and the number of new cases.

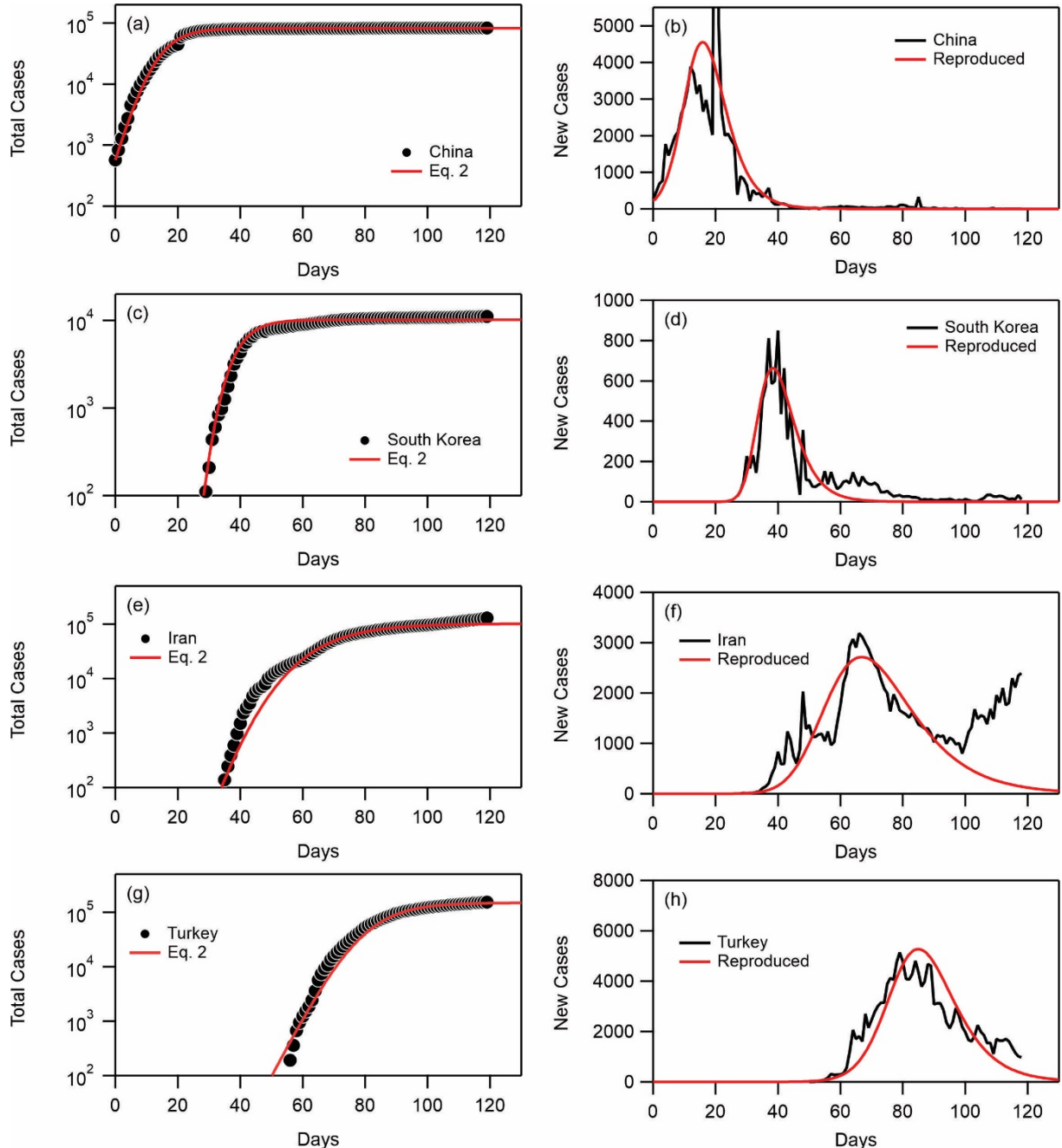

***Figure 3***: *(a, c, e, g) Total number of cases as a function of number of days starting from 22-01-2020 together with fits to Eq. 2. (a) China, (c) South Korea, (e) Iran, and (g) Turkey. The fit parameters are given in Table 2. The time has been shifted by a number of days $t_o$ for each country such that $n(t_o) = n_o$. (b, d, f, h) Number of new cases as a function of number of days starting from 22-01-2020 together with reproduced curves from the derivative of Eq. 2 using the fitted values given in Table 2. (b) China, (d) South Korea, (f) Iran, and (h) Turkey. All data are collected until 20-05-2020.*

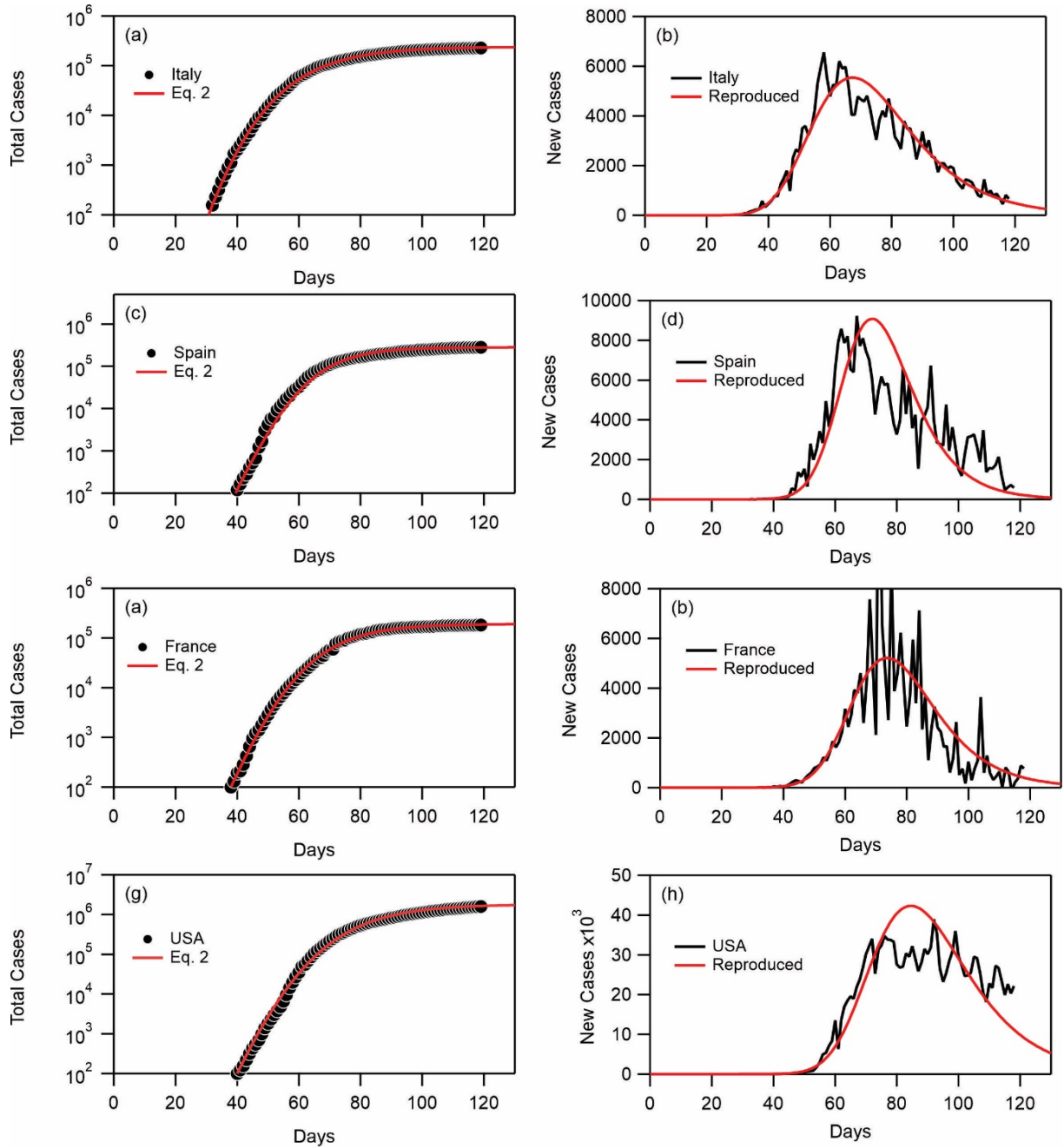

***Figure 4***: *(a, c, e, g) Total number of cases as a function of number of days starting from 22-01-2020 together with fits to Eq. 2. (a) Italy, (c) Spain, (e) France, and (g) USA. The fit parameters are given in Table 2. The time has been shifted by a number of days $t_o$ for each country such that $n(t_o) = n_o$. (b, d, f, h) Number of new cases as a function of number of days starting from 22-01-2020 together with reproduced curves from the derivative of Eq. 2 using the fitted values given in Table 2. (b) Italy, (d) Spain, (f) France, and (h) USA. All data are collected until 20-05-2020.*

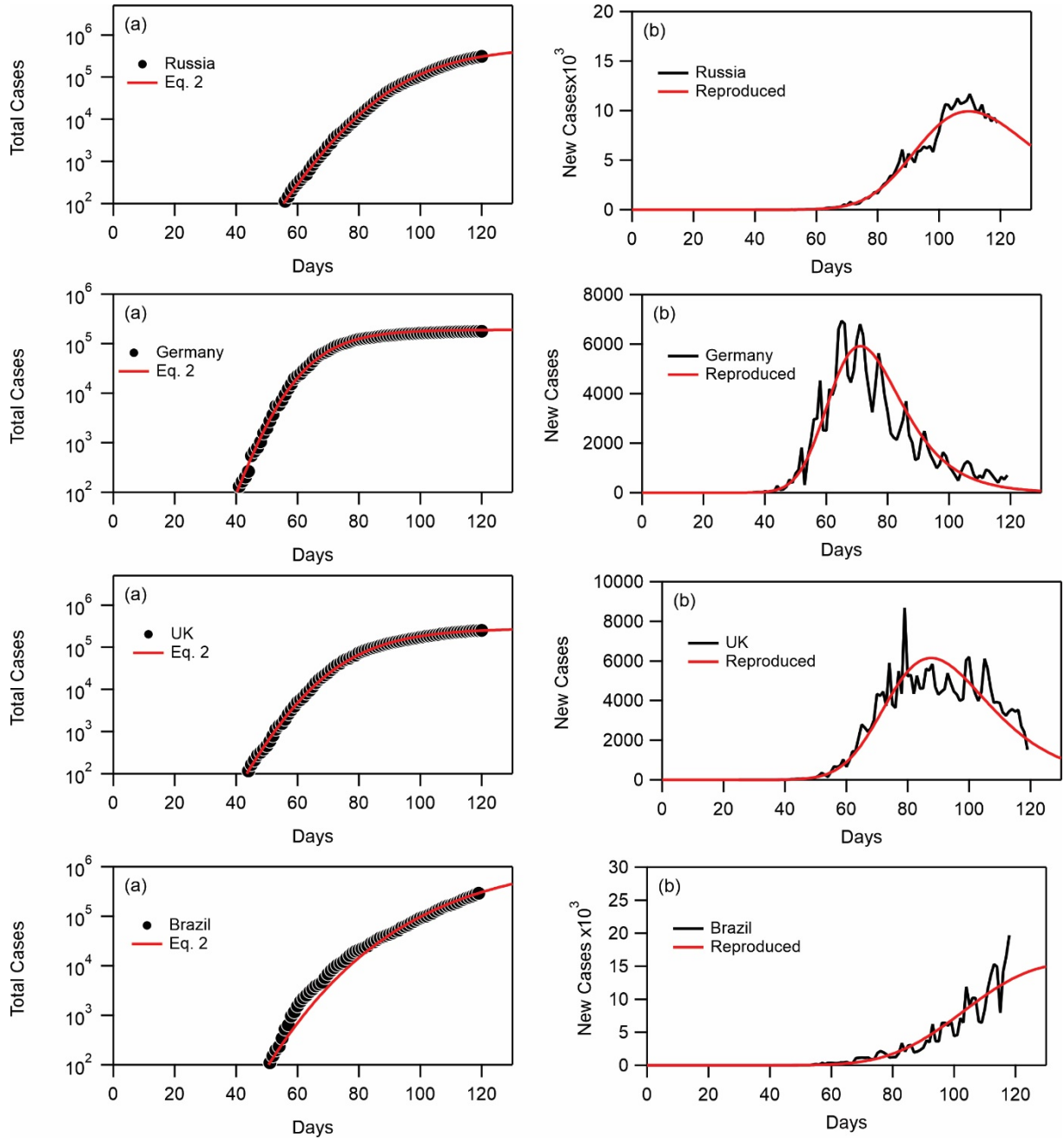

**Figure 5**: *(a, c, e, g) Total number of cases as a function of number of days starting from 22-01-2020 together with fits to Eq. 2. (a) Russia, (c) Germany, (e) UK, and (g) Brazil. The fit parameters are given in Table 2. The time has been shifted by a number of days $t_o$ for each country such that $n(t_o) = n_o$. (b, d, f, h) Number of new cases as a function of number of days starting from 22-01-2020 together with reproduced curves from the derivative of Eq. 2 using the fitted values given in Table 2. (b) Russia, (d) Germany, (f) UK, and (h) Brazil. All data are collected until 20-05-2020.*

The resulting fit parameters are summarized in Table 2. One finds that the duplication time $\tau$ varies from $1-3$ days. The significance of this range is that at the start of the epidemic, every $1-3$ days the total number of cases is almost duplicated. This observation coincides with the strong dependence of the number of new cases on the total number of cases in the $f$ phase discussed earlier. A high value of $\alpha$ parameter implies that a turnover from the $f$ phase to the $s$ phase and the $p$ phase occurs over a shorter time period as is the case for China (Table 2 & Fig. 1). For Spain, which has about the same duplication time as China, a lower value of the $\alpha$ parameter has a reverse effect. It is noted that the fitted $n_\infty$ values for all countries except China and South Korea may be used as predictions since these countries have not yet reached their $p$ phases.

**Table 2:** *Best fit parameters obtained from fits of Eq. 2 to total number of cases data of several countries. In all fits $n_o \geq 100$. Error of fit for $n_\infty$, $\tau$ and $\alpha$ are respectively $\pm 10^4$ cases, $\pm 0.1$ days, and $\pm 0.01$.*

| Country | $n_\infty \times 10^5$ (number) | $\tau$ (days) | $\alpha$ | $t_{Max}^{**}$ (days) |
|---------|---------|---------|---------|---------|
| China | 0.83 | 1.7 | 0.45 | 16 |
| South Korea | 0.10 | 0.6 | 0.16 | 8 |
| Iran | 1.02 | 1.5 | 0.17 | 32 |
| Turkey | 1.49 | 2.7 | 0.46 | 27 |
| Italy | 2.40 | 0.9 | 0.08 | 37 |
| Spain | 2.80 | 1.8 | 0.25 | 33 |
| France | 1.91 | 1.8 | 0.21 | 36 |
| USA | 18.0 | 1.4 | 0.14 | 44 |
| Russia | 5.16 | 2.4 | 0.20 | 53 |
| Germany | 1.90 | 1.5 | 0.20 | 30 |
| UK | 2.80 | 1.9 | 0.18 | 43 |
| Brazil | 14.9 | 1.0* | 0.04 | 88 |

*Fixed parameter in the fit.* **Counts from the start of the epidemic in each state, and not from 22-01-2020.*

The form of Eq. 2 may then be used to give general trends of the epidemic using the fitted values in Table 2. To this end, the number of cases at the two extremities were selected to be $n_o = 10^2$ and $n_\infty = 10^6$, the duplication time was varied in the range $\tau = 1 - 3$ days and $\alpha = 0.1 - 0.5$. In the Supporting Information Fig. SI 1, several predictions are plotted using these values. It is shown that for a fix value of $\alpha$ parameter, a higher value of the duplication time $\tau$ broadens the curves whereby the turnover period to the $s$ phase and the $p$ phase is also prolonged. Similarly, the onset of the maximum in the number of new cases is delayed. It is otherwise found that the $\alpha$ parameter has a reverse effect. For example, at a fix value of the duplication time $\tau$, a higher value of the $\alpha$ parameter shifts the curves to the left whereby the $s$ phase and the $p$ phase occur earlier, and similarly the onset of the maximum in the number of new cases is shortened. Thereby the $\alpha$ parameter, which is strictly responsible for deviation from an exponential growth, gives approximately the time range between the initial growth phases and the final $p$ phase. This variation may be abrupt when $\alpha$ is large (*e.g.* this is the case in the cell model) or may be smooth when $\alpha$ is small. These curves suggest that both $\alpha$ and $\tau$ control the trend behavior from $n_o$ to $n_\infty$. These two parameters can then be used for prediction purposes and planning regarding the trends in the other countries.

Furthermore, from the estimated values of $\tau$, $\alpha$ and $n_\infty$ in Table 2, it is possible to estimate the duration from the start of the epidemic to the onset of the maximum in the number of new cases. This duration which is denoted $t_{Max}$ is expressed as follows:

$$t_{Max} = \frac{\tau}{\alpha \ln 2} \ln \left\{ \frac{1}{\alpha} \left[ \left( \frac{n_\infty}{n_o} \right)^\alpha - 1 \right] \right\}. \tag{3}$$

The form of Eq. 3 shows that this time period is proportional to the duplication time $\tau$; however, it is inversely related to the $\alpha$ parameter. This relation agrees with our previous statement on the general trend behaviors. Furthermore, Eq. 3 shows that the dependence on the ratio of the total number of cases to the initial number of cases, $n_\infty/n_o$ is weak. The estimated $t_{Max}$ values are provided in Table 2. It is clear that China and South Korea have reached to this maximum over a shorter period of time as compared to the other countries. However, in general, $2 - 8$ weeks is the period over which these countries have reached the maximum of the number of their new cases. The onset of this maximum and the number of new cases at this maximum, which is generally compared with the medical capacity, are important parameters for planning and placing mitigating measures by the states.

## 3. Conclusions

It is concluded by the analysis presented in this letter that the first period of the Covid-19 epidemic, before the herd immunity, may be interpreted in terms of simple models containing physically meaningful parameters. The temporal evolution of the total number of cases may be divided to three phases. In the $f$ phase, it is shown that the number of new cases depends on the total number of cases in a power-law relation with an exponent $\beta = 0.82$. The overall temporal evolution of the total number of cases, *i.e.* all three phases, may also be interpreted by modeling. It is found that the duplication time $\tau$ is $1 - 3$ days and the $\alpha$ parameter, which controls the exponential growth, to be in the range $0.1 - 0.5$ for selected countries. The higher is the value of this parameter, the shorter is the period of phase change. The duplication time together with the $\alpha$ parameter may be used to interpret the total number of case as well as the number of new cases. The models presented in this work, and the average values which are obtained by fitting to several countries with leading number of cases, may be used for prediction purposes as well as for guiding the implementation of proper controlling measures during each phase of the epidemic. The presented model and analysis are for when the total number of infected cases is less than 10% of the entire population. In order to interpret the trends when up to 60% of the population are infected, a different model is required, which is out of the scope of the present work.

**Supporting Information**

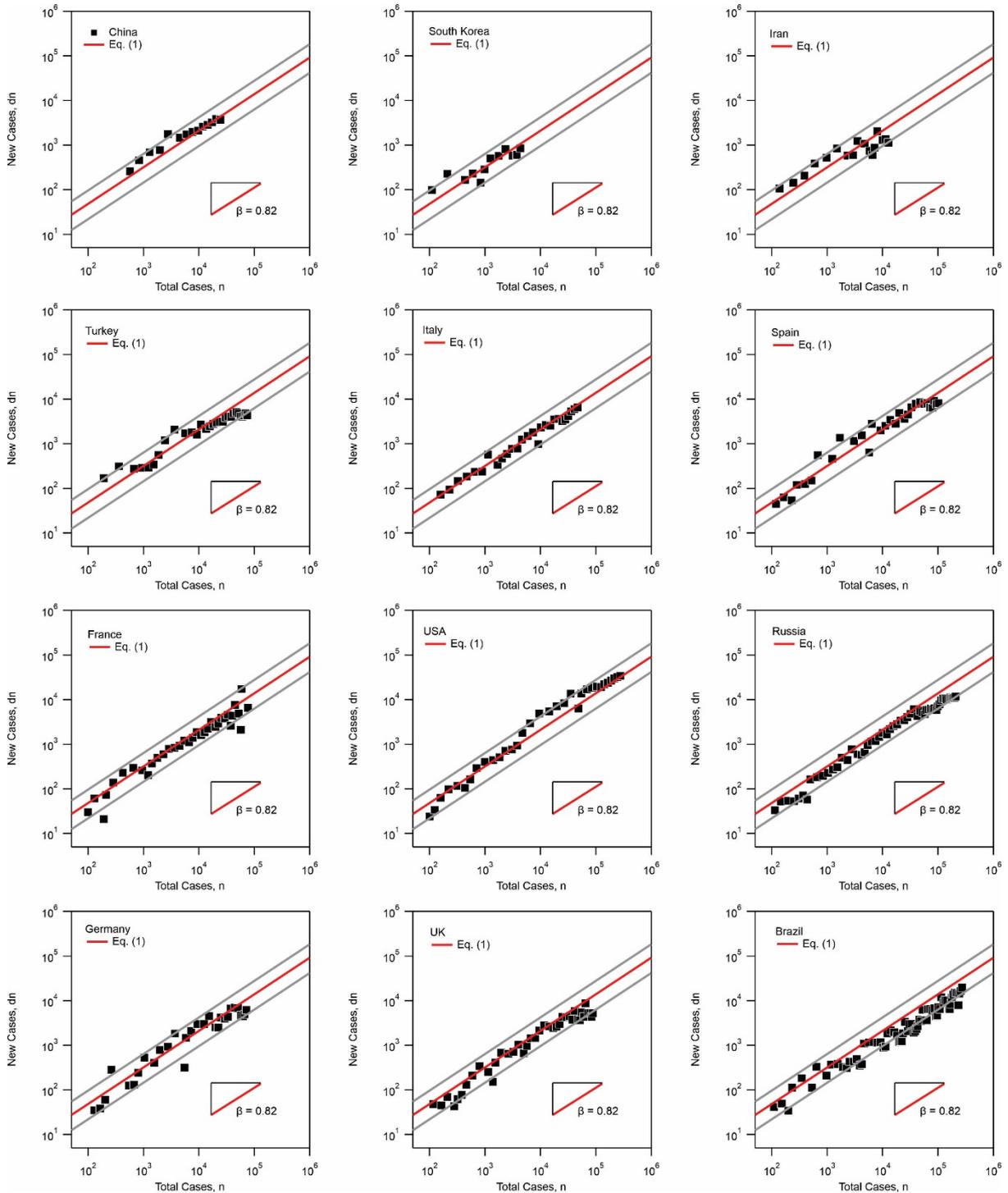

***Figure SI 1****: Number of new cases as a function of total number of cases showing the power-law dependence of Eq. 1. The exponent $\beta = 0.82$ is found to interpret the data reasonably. The prefactor $\lambda = 0.5$ gives the lower bound and $\lambda = 2.2$ the upper bound. Top to bottom and left to right: China, South Korea, Iran, Turkey, Italy, Spain, France, United States, Russia, Germany, United Kingdom, and Brazil.*

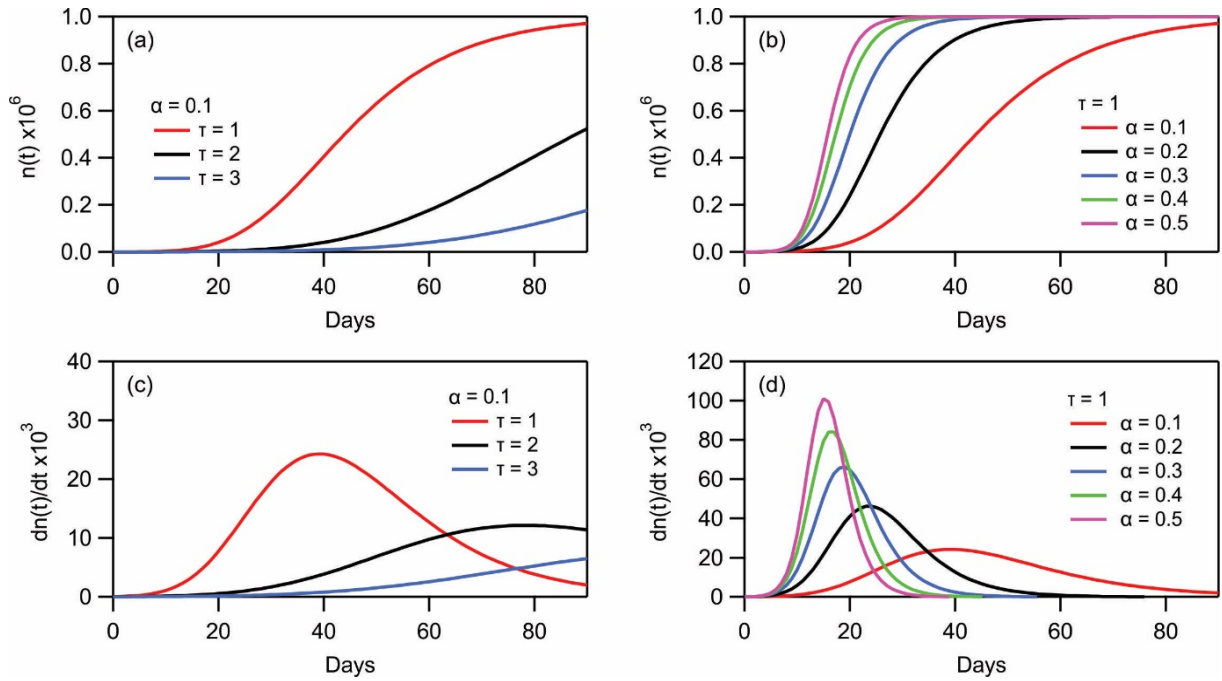

**Figure SI 2**: *Calculation of the total number of cases (a, b) and the number of new cases (c, d) from Eq. (2) in the main text. The number of new cases is equivalent to the derivative of Eq. (2) in the main text. $n_o = 10^2$, $n_\infty = 10^6$, $\tau = 1 - 3$ days, and $\alpha = 0.1 - 0.5$.*